\documentstyle[aps,12pt,preprint,psfig]{revtex}

\def \to {\rightarrow}
\def \beq {\begin{equation}}
\def \eeq {\end{equation}}
\def \ba {\begin{eqnarray}}
\def \ea {\end{eqnarray}}
\def \jpsi {J/\psi}

\def \< {\left <}
\def \> {\right >}

\begin{document}
\tightenlines
\preprint{
\hbox{PKU-TP-98-51}}
\draft

\title{Diffractive charm jet production at
        hadron colliders in the two-gluon exchange model}

\author{Feng Yuan}
\address{\small {\it Department of Physics, Peking University, Beijing 100871, People's Republic
of China}}
\author{Kuang-Ta Chao}
\address{\small {\it China Center of Advanced Science and Technology (World Laboratory), Beijing 100080,
        People's Republic of China\\
      and Department of Physics, Peking University, Beijing 100871, People's Republic of China}}

\maketitle
\begin{abstract}

We present a calculation of diffractive charm jet production at hadron
colliders in perturbative QCD based on the two-gluon exchange model.
Differing from the previous calculations, we abandon the use of the effective
color-singlet vitual gluon simplification, and use the real gluon in the
calculations for the partonic process.
We find the final result is free of the linear singularities due to
the small transverse momenta of the exchanged two gluons.
In the leading logarithmic approximation (LLA)
in QCD, this process is related to the off-diagonal gluon
density in the proton.
As a result, this process may provide a wide window
for testing the two-gluon exchange model, and may be particularly 
useful in studying the small $x$ physics.
In comparison, the diffractive bottom jet production is also
discussed, and is found to be important only for large transverse momentum
of the jet.

\end{abstract}


\section{Introduction}

In recent years, there has been a renaissance of interest in
diffractive scattering.
These diffractive processes are described by the Regge theory in
terms of the Pomeron ($I\!\!P$) exchange\cite{pomeron}.
The Pomeron carries quantum numbers of the vacuum, so it is a colorless entity
in QCD language, which may lead to the ``rapidity gap" events in experiments.
However, the nature of the Pomeron and its interaction with hadrons remain a mystery.
For a long time it had been understood that the dynamics of the
``soft Pomeron'' was deeply tied to confinement.
However, it has been realized now that much can be learned about
QCD from the wide variety of small-$x$ and hard diffractive processes,
which are now under study experimentally.
Of all these processes, the diffractive heavy quark and quarkonium production
have drawn specially attention, because their large masses provide a natural
scale to guarantee the application of perturbative QCD\cite{th1,th2,th3}.
In the framework of perturbative QCD the Pomeron is assumed to be
represented by a pair of gluons in the color-singlet state.
An important feature of this perturbative QCD model prediction is that
the cross section for the diffractive processes is expressed
in terms of the off-diagonal gluon distribution in the proton\cite{offd}.

So far the previous studies are focused on the diffractive processes
at $ep$ collider (photoproduction and DIS processes),
we should expect that the two-gluon exchange model can also be used
to describe the diffractive processes at hadron colliders.
Recently, we have extended the idea of perturbative QCD description of
diffractive processes from $ep$ colliders to hadron colliders\cite{th4},
in which the hadronic diffractive $J/\psi$ production is calculated
in the two-gluon exchange model.
In this paper, we will calculate the diffractive charm jet production
at hadron colliders in this two-gluon exchange model.
For a theoretical point of view the study of diffractive charm jet production
has some advantages compared to $\jpsi$ production, because it advoids
the ambiguities associated with the color-octet production matrix
elements\cite{nrqcd,th4} and retains the sensitivity to the off-diagonal
gluon density in the proton.

However, there exist nonfactorization effects in the hard
diffractive processes at hadron collisions\cite{preqcd,collins,soper,tev}.
These effects mainly focus on the following
two sides.
One is the so-called spectator effects\cite{soper}.
The interaction with the spectator quarks can change the probability
of the diffractive hadron emerging from collisions intact. This
implies that the extra interaction with spectators makes it less likely
for the diffractive hadron to survive\cite{survive}.
Typically, for the diffractive processes at the Tevatron, this survival
probability is shown to be about $0.1$\cite{soper,tev}.
The other one is the ``coherent diffractive" processes, in which the whole
Pomeron is involved in the hard process.
These coherent diffractive processes have been proved
to break the factorization assumption of \cite{is}.

As shown in Fig.1, the diffractive charm jet production process $p\bar p\to c\bar c p$
calculated in this paper belongs to these coherent diffractive processes.
The whole Pomeron represented by the color-singlet two-gluon system
emitted from one hadron interacts
with another hadron to produce the charm jets.
In the leading order perturbative QCD, the partonic process $gp\to c\bar c p$
is plotted in Fig.2.
In the diffractive final states, there are only charm and anticharm jets
balanced in the transverse momentum distribution.
Experimentally, this would provide a strong signal for the coherent
diffractive processes at hadron colliders.

The diffractive production of heavy quark jet at hadron colliders has also
been studied in Ref.\cite{levin}. However, our calculation
is quite different from theirs.
In their calculation of the partonic process $gp\rightarrow c\bar c p$,
they used an effective color-singlet virtual gluon (as a color-singlet
probe in their language)
to replace the real gluon for simplification.
In our calculations, we abandon this simplification, and use the real gluon
in the calculations of the partonic process.
This of course will cause more complicated calculations.
However, after a lengthy calculation which will be shown in the following
we find that our result is free of linear singularities,
and then guarantees the gauge invariance of QCD.

Another important difference between \cite{levin} and our calculations is the
definition of the coherent diffraction. Following Ref.\cite{collins}, we call
the process in which the whole Pomeron participants in the hard scattering
process as the coherent diffractive process. Under this definition, all of
the nine diagrams of Fig.2 contribute to
the coherent diffractive production of heavy quark jet. However, in \cite{levin}
they separate their diagrams into two parts: one part (the first four
diagrams) contributes to
the conventional IS model diffraction; and the other part contributes to the
so-called coherent diffraction.
However, in our calculations, the first four diagrams in Fig.2
alone do not contribute a gauge invariant part of the amplitude, and
their sum will lead to a linear singularity which is not proper in QCD
calculations.
So, for a full calculation, the nine diagrams must be summed together to
contribute to the coherent diffraction.

Another important issue of this process is
about the gluon (off-diagonal) distribution in the proton at small $x$.
As shown in Ref.~\cite{th4}, in hadron collisions (such as at the Tevatron),
we can explore the gluon distribution at small $x$ down to $10^{-6}$ by
studying the hard diffractive $\jpsi$ production.
In the diffractive charm jet production, the involved gluon distribution
function may take its value at the same order as that for $\jpsi$ production.
So, detecting the diffractive charm jet production at hadron colliders
is also attractive for the study of the small $x$ physics.
Furthermore, experimentally the diffractive charm jet production is 
easily detected because the diffractive final states of $c$ and $\bar c$ pair
may have large transverse momenta distribution.

The rest of the paper is organized as follows.
In Sec.II, we give the cross section formula for the partonic process
$gp\to c\bar c p$ in the leading logarithmic approximation (LLA) QCD.
We use the Feynman rule method in the calculations.
We also check our method to reproduce the formula for the diffractive
charm jet production in the photoproduction processes previous calculated
by others\cite{th3}.
The numerical results are given in Sec.III for diffractive charm jet
production at the Fermilab Tevatron. In Sec.IV, we give some discussions about
the off-diagonal distribution effects on the process calculated here
and the $\jpsi$ production of \cite{th4}.
The conclusion is given in Sec.V.

\section{ LLA formula for the partonic process}

As sketched in Fig.1, the cross section for the diffractive charm jet production
at hadron colliders ($p\bar p$ at the Tevatron) can be formulated as,
\beq
d\sigma(p\bar p\to c\bar cp)=\int dx_1d\hat{\sigma}(gp\to c\bar c p)g(x_1,Q^2),
\eeq
where $x_1$ is the longitudinal momentum fraction of the antiproton
carried by the incident gluon.
$g(x_1,Q^2)$ is the gluon density in the antiproton, and 
$Q^2$ is the scale of the hard process.
$d\hat{\sigma}(gp\to c\bar c p)$ is the cross section for the partonic process
$gp\to c\bar c p$.

In this section, we first give the formula for the partonic process $gp\to
c\bar c p$ in the LLA QCD. In the leading order of perturbative QCD, there
are nine diagrams shown in Fig.2, contributing to the partonic process
$gp\to c\bar c p$.
The two-gluon system coupled to the proton (antiproton) in Fig.2 is in
a color-singlet state, which characterizes the diffractive processes in
perturbative QCD.
The two-gluon exchange model itself may not account for the full structure
of the Pomeron, but it can approximate the description of diffractive processes
in terms of perturbative QCD.
Due to the positive signature of these diagrams (color-singlet exchange),
we know that the real part of the amplitude cancels out in the leading
logarithmic approximation.
To evaluate the imaginary part of the amplitude, we must calculate the
discontinuity represented by the crosses in each diagram of Fig.2.

The first four diagrams of Fig.2 are the same as those calculated in the
diffractive photoproduction processes. But, due to the existence of
gluon-gluon interaction vertex in QCD, in the partonic process
$gp\to c\bar c p$, there are additional five diagrams (Fig.2(5)-(9)).
These five diagrams are needed for complete calculations in this order of QCD.
From the following calculations, we can see that these five diagrams are
important to cancel out the anormalous singularity which rises from the first
four diagrams to obtain the correct results.

To calculate the imaginary part of the amplitude ${\cal A}(gp\to c\bar c +p)$,
we employ the Feynman rule method\cite{wust}.
For a cross check, we also use this method to calculate the
diffractive charm jet photoproduction process $\gamma p\to c\bar c p$.
In the leading logarithmic approximation, we can reproduce
the results obtained previously by using the light-cone
wave function method \cite{th3}.

\subsection{Kinematics and Sudakov variables}

In the diffractive process $gp\to c\bar c p$, the final state only contains
the charm quark and anti-charm quark. We set $\vec{k}_T$ to be the
transverse momentum of the charm quark, $m_c$ the charm quark mass, and
we define $m_T^2=m_c^2+k_T^2$.
The invariant mass of the $c\bar c$ pair of the diffractive system is
set to be $M_X^2$, and $x_{I\!\! P}$ is,
\beq
x_{I\!\! P}=\frac{M_X^2}{s},
\eeq
where $s$ is the total c.m. energy of the gluon-proton system.

In our calculations, we express the formulas in terms of the Sudakov
variables.
That is, every four-momenta $k_i$ are decomposed as,
\beq
k_i=\alpha_i q+\beta_i p+\vec{k}_{iT},
\eeq
where $q$ and $p$ are the momenta of the incident gluon and the proton,
$q^2=0$, $p^2=0$, and $2p\cdot q=W^2=s$.
$\alpha_i$ and $\beta_i$ are the momentum fractions of $q$ and $p$
respectively.
$k_{iT}$ is the transverse momentum, which satisfies
\beq
k_{iT}\cdot q=0,~~~
k_{iT}\cdot p=0.
\eeq

All of the Sudakov variables for every momentum
are determined by using the on-shell conditions
of the momenta of the external particles and the crossed lines in the diagram.

In the following, we calculate the differential cross section $d\hat{\sigma}/dt$ at
$t=0$. Namely, we set the momentum transfer squared of the diffractive process
$gp\to c\bar c p$ equal to zero, i.e., $u^2=t=0$.
$(q+u)$ is the momentum of the diffractive final state (contains charm and
anticharm jets), and then
\beq
(q+u)^2=M_X^2.
\eeq
With this equation, and the on-shell conditions of the
external proton lines,
\beq
(p-u)^2=p^2=0,
\eeq
we can determine the Sudakov variables associated with $u$ as
\beq
\alpha_u=0,~~~\beta_u=\frac{M_X^2}{s},~~~\vec{u}_T^2=0.
\eeq
In the diffractive region at hadron collisions, we know that
$M_X^2\ll s$, i.e., $\beta_u\ll 1$. So, in the following calculations,
we set $\beta_u$ to be a small parameter, and take the leading order
contributions,
and neglect the terms proportional to $\beta_u=\frac{M_X^2}{s}$.

$\alpha_k$ and $\beta_k$ are determined from the on-shell conditions of
the out-going charm quark and anticharm quark,
\ba
\nonumber
(k+q)^2&=&m_c^2,\\
(u-k)^2&=&m_c^2.
\ea
By solving the above equations, we can obtain,
\ba
\nonumber
\alpha_k(1+\alpha_k)&=&-\frac{m_T^2}{M_X^2},\\
\beta_k=-\alpha_k \beta_u&=&-\frac{M_X^2}{s}\alpha_k.
\ea
Because $\alpha_k$ is of order of $1$, the value of $m_T^2$ is the same order
as $M_X^2$.
From the last equation, we can see that $\beta_k$ is much smaller
than $\alpha_k$, i.e., $|\beta_k|\ll |\alpha_k|$,
because $\beta_u\ll 1$.

For the loop momentum $l$, we know that the integral of the amplitude
over $l_T$ receives
large logarithmic contribution from the region $1/R_N^2 \ll l_T^2 \ll M_X^2$
($R_N$ is the nucleon radius)\cite{th1}.
That is to say, the dominant contribution of the integration of $l_T$
comes from the region $l_T^2 \ll M_X^2$, so $l_T^2$ is a small parameter
compared with $M_X^2$ and $m_T^2$.
To calculate the integration of the amplitude over $l_T^2$, we can 
expand the amplitude in terms of $l_T^2$ and take the leading order
contributions.

The Sudakov variable $\alpha_l$ can be determined from the on-shell
condition of the bottom cross on the proton line in each diagram of Fig.2,
i.e.,
\beq
(p-l-u)^2=0,
\eeq
which results in
\beq
\alpha_l=-\frac{l_T^2}{s}.
\eeq

$\beta_l$ is determined from the on-shell condition of the up cross
on the charm quark line or the gluon line in each diagram.
Unlike other variables calculated above 
the value of $\beta_l$ is not the same for these nine diagrams of Fig.2, because 
there are three different on-shell conditions of the up crossed lines,
\ba
(k-l-u)^2&=&m_c^2,~~~{\rm for~Diag.}1,~3,~5,\\
(k+q+l)^2&=&m_c^2,~~~{\rm for~Diag.}2,~4,~6,\\
(q+l+u)^2&=&0,~~~~~{\rm for~Diag.}7,~8,~9.
\ea
These three different relations lead to three different values for $\beta_l$,
\ba
\beta_l&=&\frac{2(k_T,l_T)-l_T^2}{\alpha_ks},~~~{\rm for~Diag.}1,~3,~5,\\
\beta_l&=&\frac{2(k_T,l_T)+l_T^2}{(1+\alpha_k)s},~~~{\rm for~Diag.}2,~4,~6,\\
\beta_l&=&-\frac{M_X^2-l_T^2}{s},~~~~~~~{\rm for~Diag.}7,~8,~9.
\ea
To obtain the above results, we have used the approximations:
$\beta_u\ll 1$, and $|\beta_k|\ll |\alpha_k|$.

To end up the analysis of this subsection, we must note
that there are two small parameters in the above
calculations of the Sudakov variables,
\beq
\beta_u\ll 1,~~~~\frac{l_T^2}{m_T^2}\ll 1.
\eeq
And these two small parameters are the basic expansion parameters in the following
calculations of the amplitude for the partonic process $gp\to c\bar cp$.

\subsection{Expand the amplitude in terms of $l_T^2$}

Using the variables induced in the above, we evaluate the differential 
cross section formula for the partonic process $gp\to c\bar c p$ as,
\beq
\label{xs}
\frac{d\hat{\sigma}(gp\to c\bar cp)}{dt}|_{t=0}=\frac{dM_X^2d^2k_Td\alpha_k}{16\pi s^216\pi^3M_X^2}
        \delta(\alpha_k(1+\alpha_k)+\frac{m_T^2}{M_X^2})\sum \overline{|{\cal A}|}^2,
\eeq
where ${\cal A}$ is the amplitude of the process $gp\to c\bar cp$.
We know that the real part of the amplitude ${\cal A}$ is zero,
and the imaginary part of the amplitude ${\cal A}(gp\to c\bar cp)$ for each diagram
of Fig.2 has the following general form,
\beq
\label{ima}
{\rm Im}{\cal A}=C_F(T_{ij}^a)\int \frac{d^2l_T}{(l_T^2)^2}F\times\bar u
        _i(k+q)\Gamma_\mu v_j(u-k),
\eeq
where $C_F$ is the color factor for each diagram, and $a$ is the color index
of the incident gluon. $\Gamma_\mu$ is some $\gamma$ matrices including one
propagator. $F$ in the integral represents some other factor which is the
same for each diagram,
\beq
F=\frac{3}{2s}g_s^3f(x',x^{\prime\prime};l_T^2),
\eeq
where
\beq
\label{offd1}
f(x',x^{\prime\prime};l_T^2)=\frac{\partial G(x',x^{\prime\prime};l_T^2)}{\partial {\rm ln} l_T^2},
\eeq
where the function
$G(x',x^{\prime\prime};k_T^2)$ is the so-called
off-diagonal gluon distribution function\cite{offd}.
Here, $x'$ and $x^{\prime\prime}$ are the momentum fractions of the proton
carried by the two gluons.
It is expected that for small $x$, there is no large difference between the off-diagonal and
the usual diagonal gluon densities\cite{off-diag}.
So, in the following calculations, we estimate the production rate by
approximating the off-diagonal gluon density by 
the usual diagonal gluon density, 
$G(x',x^{\prime\prime};Q^2)\approx xg(x,Q^2)$, where $x=x_{I\!\! P}=M_X^2/s$.
Later we will discuss the off-diagonal parton distribution
function effects for the diffractive charm jet production process calculated
here and $\jpsi$ production.

The color factors $C_F$ are not the same for the nine diagrams, and they are
\ba
\label{cf}
\nonumber
C_F&=&\frac{2}{9},~~~~~~~{\rm for~ Diag.}1,~4,\\
\nonumber
C_F&=&-\frac{1}{36},~~~{\rm for~ Diag.}2,~3,\\
\nonumber
C_F&=&\frac{1}{4},~~~~~~~{\rm for~ Diag.}5,~8,\\
\nonumber
C_F&=&-\frac{1}{4},~~~~~{\rm for~ Diag.}6,~9,\\
C_F&=&-\frac{1}{2},~~~~~~~{\rm for~ Diag.}7,
\ea
respectively.

As mentioned above, to calculate the leading logarithmic results for the partonic
process $gp\to c\bar cp$, the amplitude of each diagram must be expanded in
terms of $l_T^2$. From the integral of 
Eq.~(\ref{ima}), we can see that the large logarithmic contribution comes from
the region $l_T^2\ll M_X^2$. So, the leading logarithmic contribution to
the cross section of the partonic process comes from the terms in $\Gamma_\mu$
which are proportional to $l_T^2$.
Because we are interested in the leading logarithmic results of the cross section,
we neglect the higher order terms of $l_T^2$ in $\Gamma_\mu$.

Futhermore, in the integral of Eq.~({\ref{ima}) the $l_T^0$ terms in $\Gamma_\mu$
coming from all diagrams must be canceled out by each other.
Otherwise, their net sum (order of $l_T^0$) will lead to a linear singularity
when we perform the integration over $l_T^2$.
The linear singularity is not proper in QCD calculations.
So, we first observe the amplitude behavior at the order of $l_T^0$, i.e., in
the limit of $l_T^2\to 0$.
In this limit, the $\Gamma_\mu$ for each diagram has the following
result,
\ba
\nonumber
\Gamma_\mu^{(1)}&=&-\Gamma_\mu^{(3)}=-\Gamma_\mu^{(5)}=\frac{s^2}{M_X^2}\alpha_k\gamma_\mu,\\
\nonumber
\Gamma_\mu^{(2)}&=&-\Gamma_\mu^{(4)}=-\Gamma_\mu^{(6)}=\frac{s^2}{M_X^2}(1+\alpha_k)\gamma_\mu,\\
\Gamma_\mu^{(7)}&=&\Gamma_\mu^{(8)}=-\Gamma_\mu^{(9)}=-\frac{s^2}{M_X^2}\gamma_\mu-\not\!p p_\mu.
\ea
The second terms in $\Gamma_\mu^{(7)}$, $\Gamma_\mu^{(8)}$ and
$\Gamma_\mu^{(9)}$ come from the contributions of $\beta_l$ terms.
This is because $\beta_l=-\frac{M_X^2}{s}\not=0$ in the limit $l_T^2\to 0$.
To obtain the above results, we have used the following identities,
\ba
\nonumber
\bar u(k+q)(\not\!k+\not\!q -m_c)&=&0,\\
\nonumber
(\not\!u-\not\!k +m_c)v(u-k)&=&0,\\
\bar u(k+q)(\not\!q+\not\!u)v(u-k)&=&\bar u [(\not\!k+\not\!q -m_c)+
        (\not\!u-\not\!k +m_c)]v=0.
\ea
With the values of the color factor $C_F$ for the diagrams given in Eq.~(\ref{cf}),
we find that their sum is zero. That is to say, there is no contribution
to the amplitude of the partonic process $gp\to c\bar cp$ in this 
order ($\Gamma_\mu$ takes the order of $l_T^0$).
As mensioned above, this is what we expected.
If there are only the first four diagrams (the same as those in photoproduction
processes), their contributions of the order of $l_T^0$ can not be canceled
out by each other, because their color factors are not the same for Diag.1,4
and Diag.2,3. (In photoproduction processes, the color factors for these
four diagrams are the same, so their contribution is zero at the order of $l_T^0$.)
However, as mentioned above, QCD gauge invariance requires the appearance of
other five diagrams at the same order of coupling
constant of strong interaction.
The contributions from Diag.5 and 6 cancel out the linear singularity
which rises from the first four diagrams.
The linear singularity from the last three diagrams, Diag.7,~8, and 9, are
canceled out by each other.
So, the amplitude has no linear singularity now.

From the above analysis, we see that the first four diagrams alone cannot give
a gauge invariant part of the amplitude for the partonic process $gp\to c\bar cp$.

At the next order expansions of $\Gamma_\mu$, $l_T^2$, the evaluation is
much more complicated. We first give the expansion results for the propagators
in $\Gamma_\mu$ for every diagrams.
To the order of $l_T^2$, the expansions of these propagators are,
\ba
\nonumber
g_1&=&\frac{1}{\alpha_kM_X^2},~~~~g_4=-\frac{1}{(1+\alpha_k)M_X^2},\\
\nonumber
g_2&=&\frac{1}{\alpha_kM_X^2}[1-\frac{2(k_T,l_T)}{m_T^2}-\frac{l_T^2}{m_T^2}+\frac{4(k_T,l_T)^2}{(m_T^2)^2}],\\
\nonumber
g_3&=&-\frac{1}{(1+\alpha_k)M_X^2}[1+\frac{2(k_T,l_T)}{m_T^2}-\frac{l_T^2}{m_T^2}+\frac{4(k_T,l_T)^2}{(m_T^2)^2}],\\
\nonumber
g_5&=&\frac{1}{M_X^2}[1-\frac{2(k_T,l_T)}{\alpha_kM_X^2}+\frac{(1+\alpha_k)l_T^2}{\alpha_kM_X^2}+\frac{4(k_T,l_T)^2}{(\alpha_kM_X^2)^2}],\\
\nonumber
g_6&=&\frac{1}{M_X^2}[1-\frac{2(k_T,l_T)}{(1+\alpha_k)M_X^2}+\frac{\alpha_kl_T^2}{(1+\alpha_k)M_X^2}+\frac{4(k_T,l_T)^2}{((1+\alpha_k)M_X^2)^2}],\\
g_7&=&\frac{1}{M_X^2},~~~~g_8=\frac{1}{\alpha_k}g_5,~~~~g_9=-\frac{1}{1+\alpha_k}g_6.
\ea

Apart from the propagator expansions, the $\gamma$ matrices in $\Gamma_\mu$
also contain the $l_T^2$ terms.
They are mostly coming from the Sudakov variables (expressed in $l_T^2$)
of momentum $l$, i.e., $\alpha_l=\frac{l_T^2}{s}$, $\beta_l$,
and the slasher $\not\! l_T$.
Furthermore, the $2$-dimensional product
$(k_T,l_T)$ also contributes $l_T^2$ terms. After integrating the
azimuth angle of $\vec{l_T}$, we will get the following results,
\ba
\nonumber
\int d^2l_T(k_T,l_T)^2&=&{\pi\over 2}\int dl_T^2k_T^2 l_T^2,\\
\int d^2l_T(k_T,l_T)\!\not l_T&=&{\pi\over 2}\int dl_T^2\!\not k_T l_T^2.
\ea

To test the validity of our procedure, we first evaluate the expansions of
$\Gamma_\mu$ for the first four diagrams, and use these expansions to
calculate the amplitude for the diffractive photoproduction processes
$\gamma p\to c\bar c p$. To do this, we only need do the following changes
in the amplitude formula Eq.~(\ref{ima}),
\beq
\label{cf2}
C_FT^a_{ij}\to \frac{2}{9}\delta_{ij}.
\eeq
To the order of $l_T^2$, the expansions of $\Gamma_\mu$ for the first four
diagrams are:
\ba
\label{f4}
\nonumber
\Gamma_\mu^{(1)}&=&\Gamma_\mu^{(4)}=\frac{l_T^2}{M_X^2}s\gamma_\mu,\\
\nonumber
\Gamma_\mu^{(2)}&=&\frac{l_T^2}{M_X^2}[-\not\!p\not\!q\gamma_\mu-\frac{1+\alpha_k}{\alpha_k}
        \gamma_\mu\not\!q \not\!p+\frac{1}{\alpha_k}\not\!p\gamma_\mu
        \not\!p+\frac{k_T^2-m_c^2}{(m_T^2)^2} s^2(1+\alpha_k)\gamma_\mu\\
\nonumber
        &~&-\frac{1}{\alpha_km_T^2}(\alpha_ks\not\!p\not\!k_T \gamma_\mu
        +(1+\alpha_k)s\gamma_\mu\not\!k_T\not\!p)] ,\\
\nonumber
\Gamma_\mu^{(3)}&=&\frac{l_T^2}{M_X^2}[-\gamma_\mu\not\!q\not\!p-\frac{\alpha_k}{1+\alpha_k}
        \not\!p \not\!q\gamma_\mu-\frac{1}{1+\alpha_k}\not\!p\gamma_\mu
        \not\!p-\frac{k_T^2-m_c^2}{(m_T^2)^2} s^2\alpha_k\gamma_\mu\\
        &~&+\frac{1}{(1+\alpha_k)m_T^2}(\alpha_ks\not\!p\not\!k_T \gamma_\mu
        +(1+\alpha_k)s\gamma_\mu\not\!k_T\not\!p)].
\ea
Using these expansions, together with Eqs. (\ref{cf2}), (\ref{xs}) and (\ref{ima}), we
obtain the cross section formula for the diffractive photoproduction
process $\gamma p \to c\bar c p$, which is consistent with the LLA result
calculated via light-cone wave function approach (for $Q^2=0$)
\cite{th3}.

For the last five diagrams, the expansions of $\Gamma_\mu$ are more
complicated. To simplify our calculations, the contributions of
these five diagrams including the color factor $C_F$
are added together, and decomposed into several terms as follows.

The terms (defined as $a$-term) coming from the slasher $\not\!l_T$ in the
$\gamma$ matrices, for all these five diagrams to sum up together, are
\beq
C_F\Gamma_\mu^{(a)}=-\frac{l_T^2}{M_X^2}(1+\frac{1}{4\alpha_k(1+\alpha_k)})
        \not \!pp_\mu.
\eeq
The terms ($b$-term) from $\alpha_l=-\frac{l_T^2}{M_X^2}$ are
\beq
C_F\Gamma_\mu^{(b)}=\frac{l_T^2}{4M_X^2}[(4+\frac{1}{\alpha_k(1+\alpha_k)})
        \beta_u\not\!pp_\mu-\frac{1+\alpha_k}{\alpha_k}\gamma_\mu\not\!q\not\!p-
        \frac{\alpha_k}{1+\alpha_k}\not\!p\not\!q\gamma_\mu-\frac{3}{2}s\gamma_\mu].
\eeq
The terms ($c$-term) from $\beta_l$ are
\beq
C_F\Gamma_\mu^{(c)}=\frac{l_T^2}{2M_X^2}\not \!pp_\mu[\frac{k_T^2}{m_T^2}
        -\frac{1}{2}(1-\frac{2k_T^2}{m_T^2})\frac{\alpha_k^2+(1+\alpha_k)^2}
        {\alpha_k(1+\alpha_k)}].
\eeq
The terms ($d$-term) from propagator expansions are
\beq
C_F\Gamma_\mu^{(d)}=\frac{l_T^2}{4(M_X^2)^2}s^2\gamma_\mu[\frac{\alpha_k^3-(1+\alpha_k)^3}
        {\alpha_k(1+\alpha_k)}+\frac{2k_T^2}{M_X^2}\frac{\alpha_k^3-(1+\alpha_k)^3}
        {\alpha_k^2(1+\alpha_k)^2}].
\eeq
The last term comes from such terms which are proportional to  ``$(k_T,l_T)\not\!l_T$".
They are ($e$-term)
\ba
\nonumber
C_F\Gamma_\mu^{(e)}&=&\frac{l_T^2}{4(M_X^2)^2}[-\frac{1+\alpha_k}{\alpha_k}
        (2\alpha_ks\not\!pk_T^\mu+s\gamma_\mu\not\!k_T\not\!p-\not\!q\not\!k_T
        \not\!pp_\mu-\alpha_ks\not\!k_Tp_\mu)\\
        &~&+\frac{\alpha_k}{1+\alpha_k}
        (2(1+\alpha_k)s\not\!pk_T^\mu-s\not\!p\not\!k_T\gamma_\mu+\not\!p\not\!k_T
        \not\!qp_\mu-(1+\alpha_k)s\not\!k_Tp_\mu)].
\ea

Adding up all of the above $a,~b,~c,~d,~e$ terms, and those terms coming from the
first four diagrams in Eq.~(\ref{f4}), we get the amplitude squared for the diffractive
process $gp\to c\bar c p$ as, averaged over the spin and color freedoms,
\beq
\sum \overline{|{\cal A}|}^2=\frac{\alpha_s^3\pi^5}{9(M_X^2)^4(m_T^2)^4}m_c^2s^2
        (m_T^2M_X^2-8k_T^2m_c^2)(10M_X^2-27m_T^2)^2(xg(x,Q^2))^2,
\eeq
where the scale $Q^2$ is set to be $Q^2=m_T^2$.
The factorization scale in the gluon density is very important because
we know that the parton distributions at small $x$ change rapidly with
this scale.
Here, we choose the scale to be $m_T^2$
which is typically used in the calculations of the heavy quark production processes.
In Ref.\cite{levin}, they used different scales for their two parts of
their diagrams for the partonic process. However, as discussed in the above calculations,
all of the nine diagrams must be summed together to give a gauge invariant
amplitude for the coherent
diffractive charm jet production at hadron colliders. So, the scales of the
nine diagrams must be the same.
The separation in \cite{levin} is not gauge invariant.

Finally, the cross section for the partonic process $gp\to c\bar c+p$, in the
LLA of perturbative QCD, is
\ba
\label{xs1}
\nonumber
\frac{d\hat{\sigma}(gp\to c\bar c+p)}{dt}|_{t=0}&=&\frac{dM_X^2dk_T^2}{(M_X^2m_T^2)^4}
        \frac{\pi^2\alpha_s^3}{128\times 9}\frac{m_c^2}{\sqrt{M_X^2(M_X^2-4m_T^2)}}\\
        &~&(M_X^2m_T^2-8k_T^2m_c^2)
        (10M_X^2-27m_T^2)^2(xg(x,Q^2))^2.
\ea
The integral region of $k_T^2$ satisfies the following constrain
\beq
\label{ktm}
k_T^2\le \frac{M_X^2}{4}-m_c^2.
\eeq
From Eq.(\ref{xs1}), we can see that the cross section of
diffractive light quark (high $k_T$) jet hadroproduction vanishes as $m_q^2$
($m_q$ is the quark mass)
in the leading logarithmic approximation of perturbative QCD (see also \cite{dijet}).

\section{Numerical results}

With the cross section formula Eq.~(\ref{xs1}) for the partonic process
$gp\to c\bar c p$, we can get the cross
section of diffractive production at the hadron level.
However, as mentioned above, there exist nonfactorization effects caused by
the spectator interactions in the hard 
diffractive processes in hadron collisions.
Here, we use a suppression factor ${\cal F}_S$ to describe this
nonfactorization effects in the hard diffractive processes at hadron
colliders\cite{soper}.
At the Tevatron,
the value of ${\cal F}_S$ may be as small as ${\cal F}_S\approx 0.1$\cite{soper,tev}.
That is to say, the total cross section of the diffractive processes
at the Tevatron may be reduced down by an order of magnitude due to
the nonfactorization effects.
In the following numerical calculations, we adopt this suppression factor value
to evaluate the diffractive production rate of charm jet at the Fermilab Tevatron.

Our numerical results are plotted in Fig.3 to Fig.7.
From Eq.~(\ref{xs1}), we can see that the cross section is sensitive
to the gluon density in the proton. In fact, the cross section for $p\bar p\to c\bar c p$
behaves as
\beq
\label{xsb}
d\sigma\propto g(x_1,Q^2)(g(x,Q^2))^2,
\eeq
where $x_1$ is the longitudinal momentum fraction of the proton (or antiproton)
carried by the incident gluon.
So, the c.m. energy of the gluon-proton
system is $s=x_1S$, where $S$ is the total c.m. energy of the proton
and proton (antiproton) system (e.g., $S=(1800GeV)^2$ at the Tevatron).
Then, $x=x_{I\!\! P}=M_X^2/s=M_X^2/x_1S$.
We can see that at the Tevatron, $x$ may be lowered down to the $10^{-6}\sim 10^{-5}$ level
(varying with the lower bound of the transverse momentum $k_T$ for the observation).

In the high energy diffractive processes, we know that the relation 
$M_X^2\ll s=x_1S$ must be satisfied (in our calculations, we set $M_X^2/s<0.1$).
Together with Eq.~(\ref{ktm}), these constrains
will give the integration bound of the arguments $x_1$, $M_X^2$, and $k_T^2$.

In our calculations, the charm quark mass is set to be $m_c=1.5GeV$,
and the scale of the running coupling constant of strong interaction
is set to be $Q^2=m_T^2$.
For the gluon distribution function, we choose the GRV NLO set
\cite{grv}.

In Fig.3, we show the $M_X^2$ dependence of the differential 
cross section $d^2\sigma/dtdM_X^2|_{t=0}$
for the diffractive charm jet production at the Tevatron.
In Fig.4, we plot the cross section $d\sigma/dt|_{t=0}$ as a function of lower bound
of the transverse momentum $k_T$.
From this figure, we can see that the cross section is sensitive
to the lower bound of $k_T$.
It decreases rapidly as $k_{T{\rm min}}$ increases.
In this figure, we also plot the cross section for bottom quark jet
diffractive production by using the same formula Eq.~(\ref{xs1})
except changing the quark mass from $m_c=1.5~GeV$ to $m_b=4.9~GeV$.
The difference of these two curves shows the dependence of the cross section
on the quark mass.
From Eq.~(\ref{xs1}), we can see that at large $k_T$ ($k_T\gg m_b$),
the cross section is proportional to the quark mass,
so the cross section of bottom jet is larger than that of charm jet.
However, due to the existence of the power factor of 
$1/m_T^2=1/(k_T^2+m_q^2)$ in the cross section formula,
at small $k_T$ smaller quark mass will result in larger cross section.
So, the cross section of charm jet is larger than that of bottom jet
at small $k_{T{\rm min}}$.

In Fig.5-7, we show the sensitivity of the cross section
to the gluon distribution in the proton. In Fig.5, we select $x_{1{\rm min}}$
as argument to show the behavior of the differential cross section $d\sigma/dt|_{t=0}$
as a function of $x_{1{\rm min}}$.
We can see that most contribution comes from $x_1> 10^{-2}$ region.
In Fig.6, we plot the cross section as a function of $x_{\rm max}$,
which shows that most contribution comes from $x\sim 10^{-3}-10^{-2}$ region.
In these two figures, for both we set the lower cut of the transverse momentum
$k_T$ to be $k_{T{\rm min}}=5~GeV$.
From Fig.4, we see that the lower bound of transverse momentum $k_{T{\rm min}}$
largely affects the total cross section of this process.
Furthermore, it is expected that $k_{T{\rm min}}$ can also affect the
relative importance of different $x$ region contribution to the total cross
section. This effect can be seen from Fig.7.
In this figure, we plot the ratio $\sigma/\sigma_{tot}$
as a function of $x_{\rm max}$, where $\sigma_{tot}$ is the total cross section
after integrating over all $x$ region,
while $\sigma$ is the cross section with the upper bound $x_{\rm max}$ for $x$.
From left to right, these
three curves correspond to three different lower
bound of transverse momentum, $k_{T{\rm min}}=0,~5,~10~GeV$ respectively.
From Fig.7, we see that for smaller value of $k_{T{\rm min}}$, the dominant
contribution comes from smaller $x$ region.

As stated in the section of Introduction, the nine diagrams in our
calculations all contribute to the coherent diffractive charm jet production at
hadron colliders. So, the above calculated numerical results from Figs.3-7 are all
for the coherent diffractive production.

\section{The off-diagonal gluon distribution effects}

The off-diagonal parton distribution function is a hot topic in recent
years\cite{ji1}.
Typically, the off-diagonal distribution has two different regions.
One is $x'>x_{I\!\! P}$ ($x''>0$, because $x'-x''=x_{I\!\! P}$), and
the other is $x'<x_{I\!\! P}$.
For the off-diagonal gluon distribution function $G(x',x'';Q^2)$,
$x'$ and $x''$ are the momentum fractions of the proton carried by
the outgoing gluon and the returning gluon respectively.
In the first region, because $x''>0$, $G(x',x'';Q^2)$ is similar to the
usual gluon distribution function $g(x;Q^2$). However,
in the second region ($x"<0$), $G(x',x'';Q^2)$ is rather a distribution amplitude
of taking two gluons out of the proton, which is similar to a meson's wave function.
Previous studies\cite{off-diag} have shown that in the first region, the
off-diagonal gluon distribution function is not very different from the usual diagonal
gluon distribution at small $x$.
So, in this region we can safely approximate the off-diagonal gluon distribution
function by the diagonal distribution.
On the other hand, in the region of $x'<x_{I\!\! P}$, where the
off-diagonal gluon distribution function
is distinctively different from the traditional diagonal gluon distribution function
\cite{martin},
the off-diagonal distribution function effects may be important.

From Fig.2, we find that there are three different values for $x'$ ($x''$)
because there are three different values for $\beta_l$,
\ba
\nonumber
x'=\beta_l+\beta_u&=&x_{I\!\! P}+\frac{2(k_T,l_T)-l_T^2}{\alpha_ks}~~{\rm for ~Diag.1,~3,~5},\\
\nonumber
  &=&x_{I\!\! P}+\frac{2(k_T,l_T)+l_T^2}{(1+\alpha_k)s}~~{\rm for~ Diag.2,~4,~6},\\
  &=&\frac{l_T^2}{s}~~~~~~~~~~~~~~~{\rm for~ Diag.7,~8,~9}.
\ea
When integrating the amplitude over $\vec{l}_T$, $x'$ may be
smaller than $x_{I\!\! P}$ in some region,
where the off-diagonal gluon diatribution function $G(x',x'';l_T^2)$
will take its value in the second region.
Especially, for Diag.7-9, the dominant contribution to the integration of the
amplitude comes from the region where 
$x'$ is mostly smaller than $x_{I\!\! P}$, because the large logarithism
contribution of the integration comes from the region of $x'\ll x_{I\!\! P}$ ($l_T^2\ll M_X^2)$.
So, we can expect that the off-diagonal effects on this process
may be important.

In our previous calculations of the diffractive $\jpsi$ production
at hadron colliders, we also neglect the off-diagonal gluon distribution
function effects. However,
from Fig.2 of Ref.~\cite{th4}, we see that $x'$ has two different values,
\ba
\nonumber
x'&=&\frac{M_X^2+2l_T^2}{s}=x_{I\!\!P}+\frac{2l_T^2}{s}, ~~~{\rm for~ Diag.} ~a,~b,~c,\\
        &=&\frac{l_T^2}{s},~~~~~~~~~~~~~~~~~~~~~~~~~~{\rm for~Diag.} ~d,~e.
\ea
When integrating the amplitude over $l_T^2$, the off-diagonal gluon distribution
functions for Diag.$a,~b,~c$ take their values in the region of $x'<x_{I\!\! P}$,
in which we can safely use the usual gluon distribution instead.
However, for Diag.$d,~e$, the off-diagonal gluon distributions dominantly
take their values in the region of $x'<x_{I\!\! P}$ (because the large logarithmic
contribution to the amplitude comes from the integral region $l_T^2\ll M_\psi^2$).
So, the off-diagonal distribution effects in the diffractive $\jpsi$ production
at hadron colliders are also expected to be very important.

To evaluate the off-diagonal distribution effects on the diffractive processes
(charm jet and $\jpsi$ production at hadron colliders) discussed above,
we must integrate the amplitude including the differential off-diagonal
gluon distribution function $f(x',x'';l_T^2)$ in Eq.~(\ref{offd1}) (with the
evolution scale at $Q^2=l_T^2$) over $l_T^2$ numerically.
This task is difficult and beyond the scope of this paper.
Further investigation in this direction is in progress.

\section{Conclusions}

In this paper, we have calculated the diffractive charm jet production
at hadron colliders in perturbative QCD by using the two-gluon exchange model.
Differing significantly from other models, our coherent diffractive amplitude
for the partonic process is gauge invariant and contains no linear singularities
as $l_T^2\to 0$.
We give the formula for the cross section of this process in the leading
logarithmic approximation.
By neglecting the off-diagonal gluon distribution effects, we have estimated the
production rate of diffractive charm jet production at the Tevatron.
The comparison of production rates between the charm jet and the bottom jet
has shown the dependence of the cross section on the quark mass,
which is a distinctive feature of these processes.
We have also discussed the off-diagonal gluon distribution effects on this
process and the hard diffractive $\jpsi$ production previously calculated
\cite{th4}.

In conclusion, we have calculated the heavy quark jet
production in the diffractive processes, which belongs to the coherent
diffractive processes at hadron colliders.
Detecting the heavy quark jet in hadron collisions would provide
strong signals for these coherent diffractive processes.
We note that the process calculated in this paper is gluon initiated process,
which is sensitive to small-$x$ gluon distribution in the proton.
Some other processes, such as $W^{\pm}$, Drell-Yan processes, are
quark initiated processes, which would be sensitive to the large-$x$ quark
distribution function in the proton\cite{collins}.
Work in this direction is in progress.

\acknowledgments
This work was supported in part by the National Natural Science Foundation
of China, the State Education Commission of China, and the State
Commission of Science and Technology of China.

% ======================================================================
% References
% ======================================================================
%\newpage

\newpage
\vskip 10mm
\centerline{\bf \large Figure Captions}
\vskip 1cm
\noindent
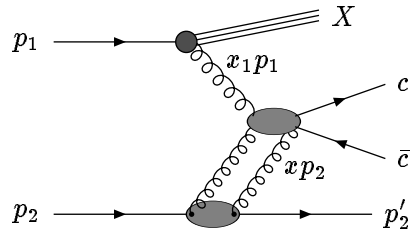
Fig.1. Sketch diagram for the diffractive charm jet production at hadron colliders
in perturbative QCD. 

\noindent
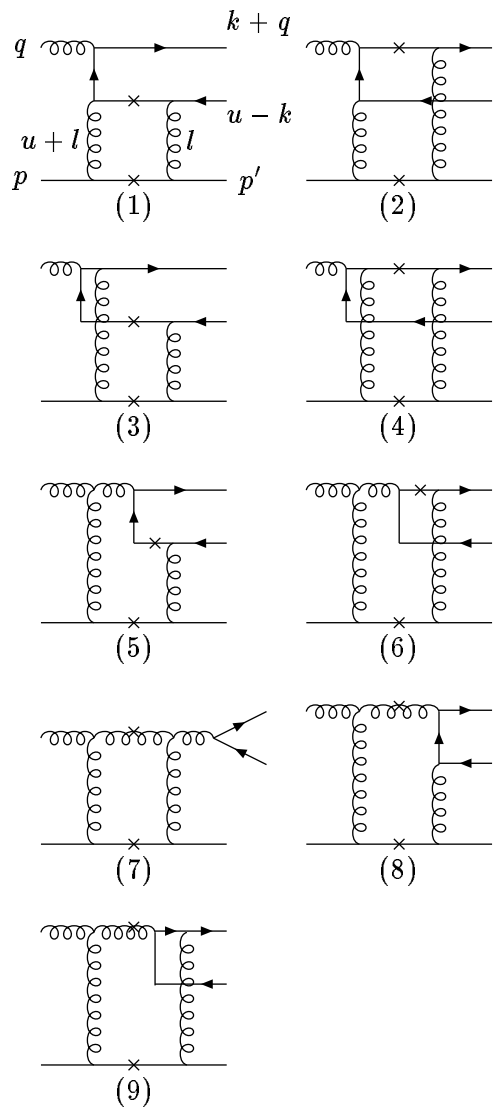
Fig.2. The lowest order perturbative QCD diagrams for partonic process
$gp\to c\bar c p$.

\noindent
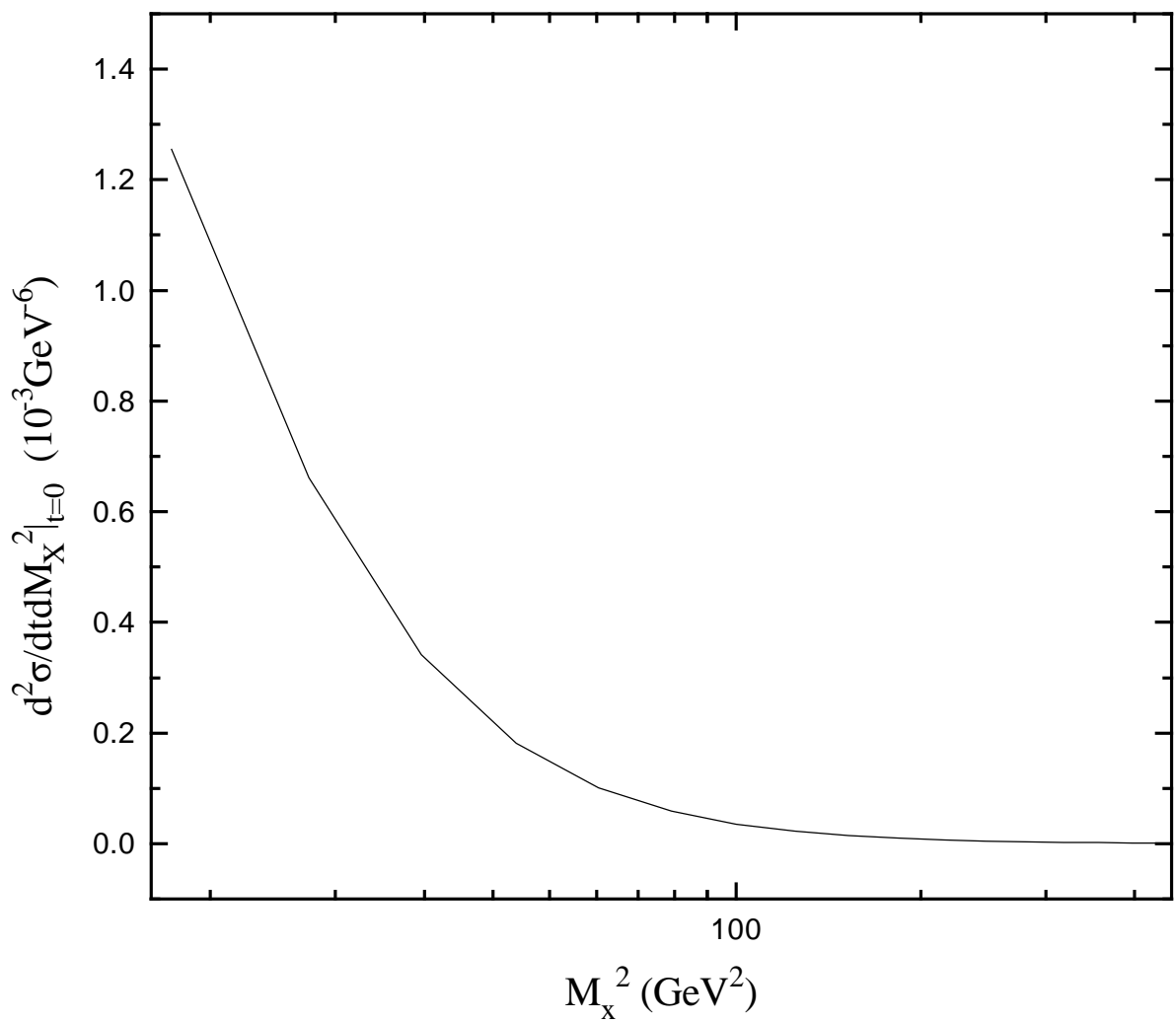
Fig.3. The differential cross section $d^2\sigma/dtdM_X^2|_{t=0}$ at the Fermilab
Tevatron as a function of $M_X^2$.

\noindent
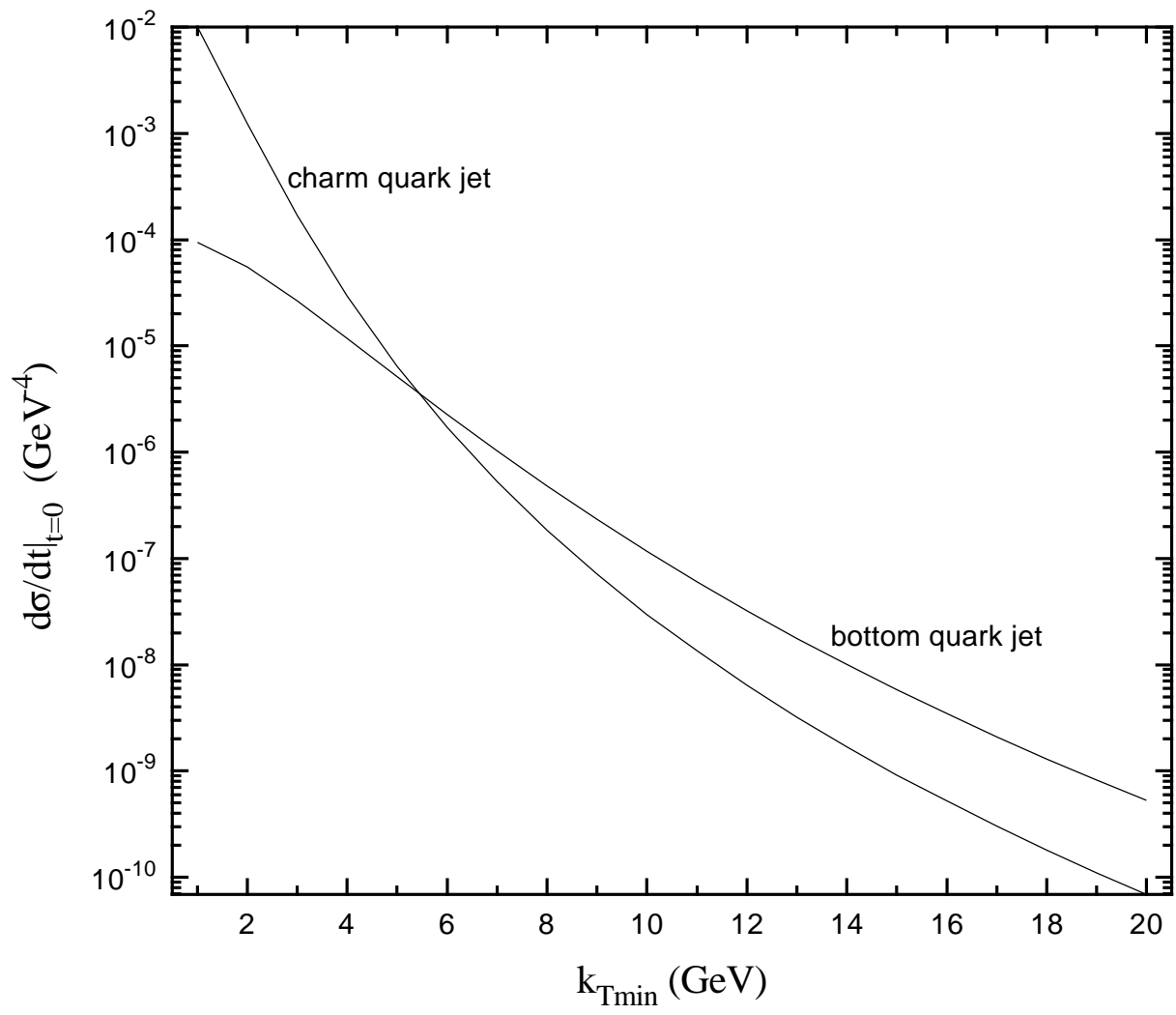
Fig.4. The differential cross section $d\sigma/dt|_{t=0}$ after integrating
over $k_T^2$ (for $k_T^2\ge k_{T{\rm min}}^2$) as a function of $k_{T{\rm min}}$,
where $m_c=1.5~GeV$ for the charm jet and $m_b=4.9~GeV$ for the bottom jet.

\noindent
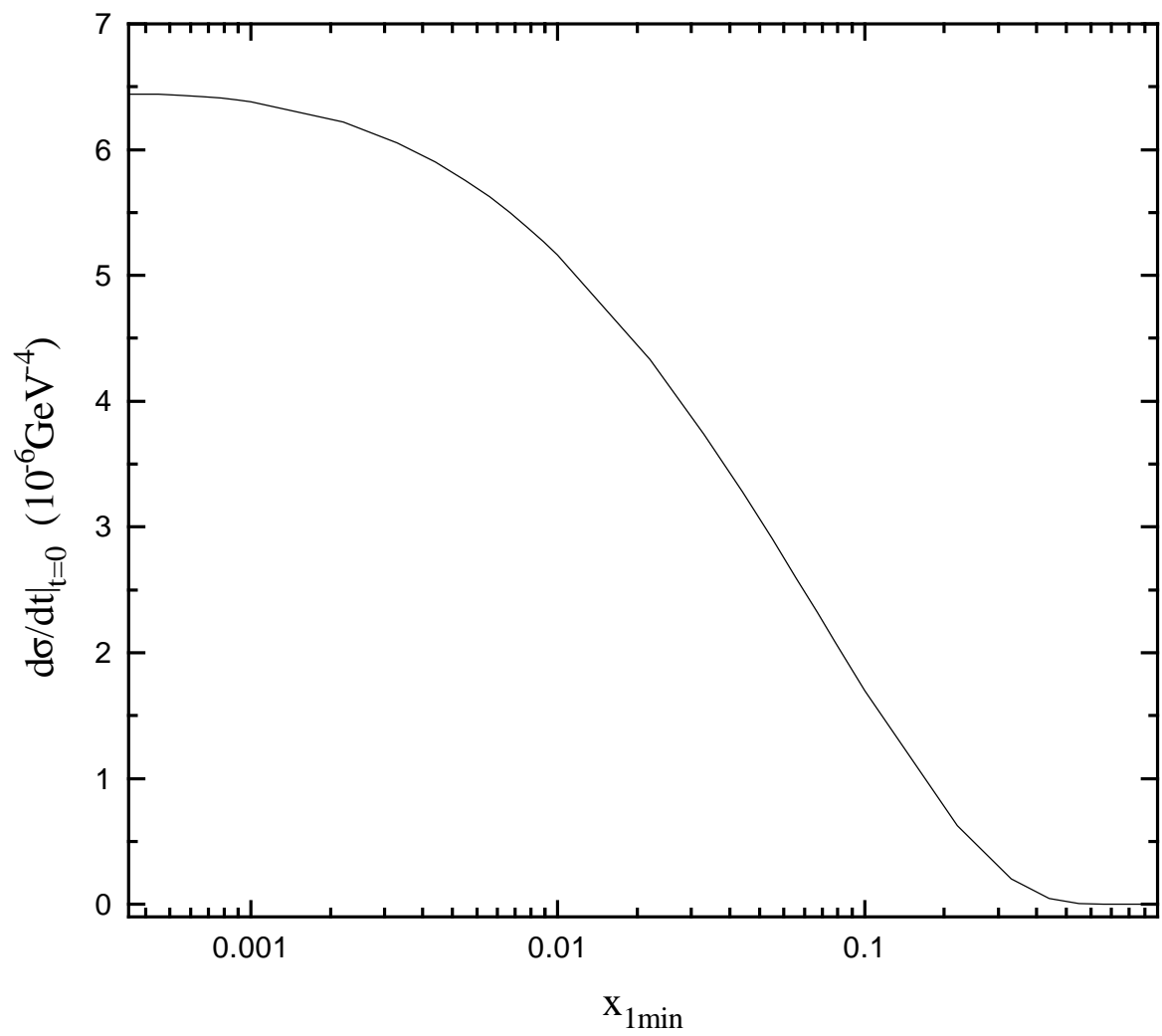
Fig.5. The differential cross section $d\sigma/dt|_{t=0}$
as a function of $x_{1{\rm min}}$, where $x_{1{\rm min}}$ is the lower
bound of $x_1$ in the integration of the cross section.

\noindent
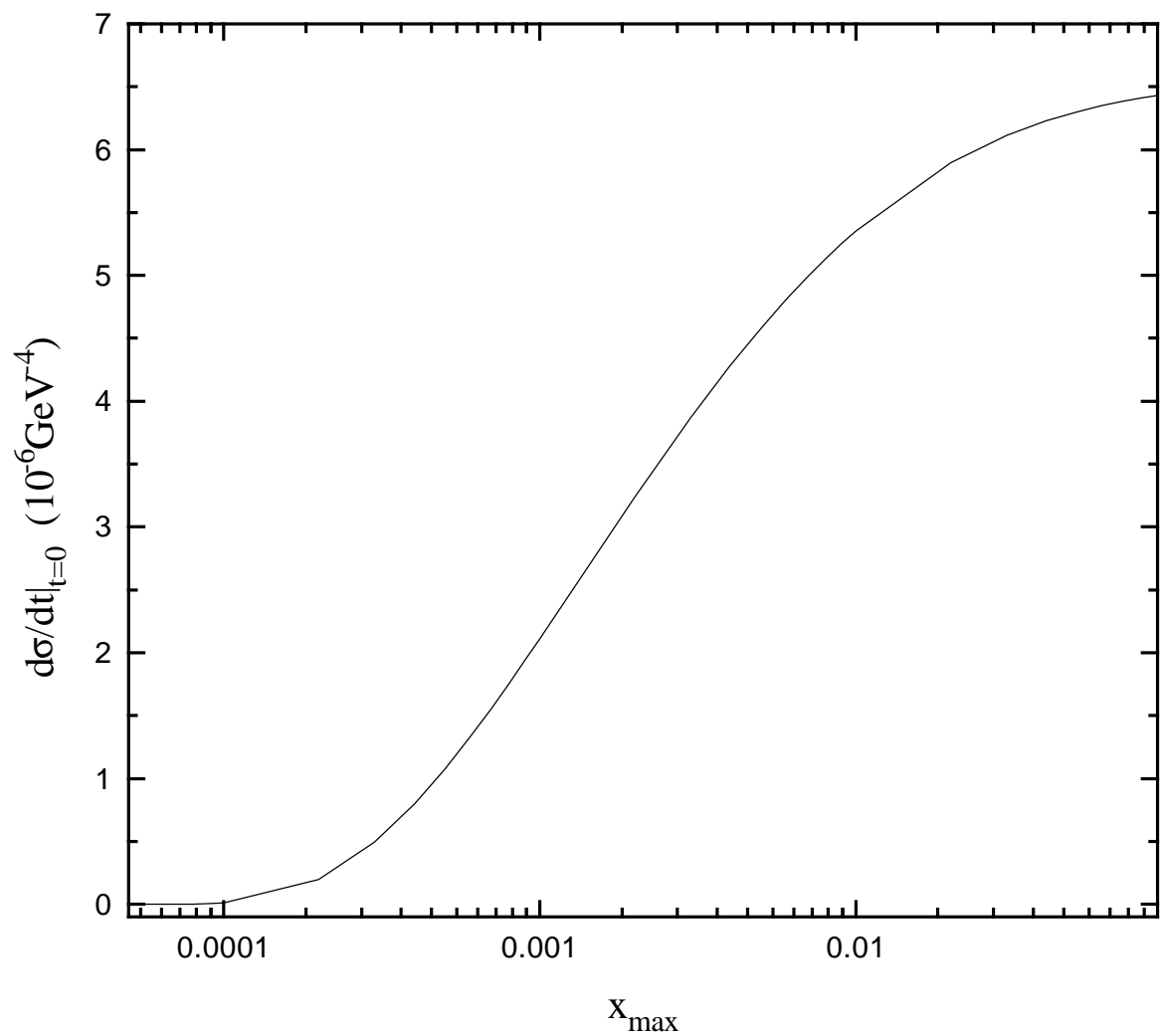
Fig.6. The differential cross section $d\sigma/dt|_{t=0}$
as a function of $x_{\rm max}$, where $x_{\rm max}$ is the upper bound
in the integration of the cross section.

\noindent
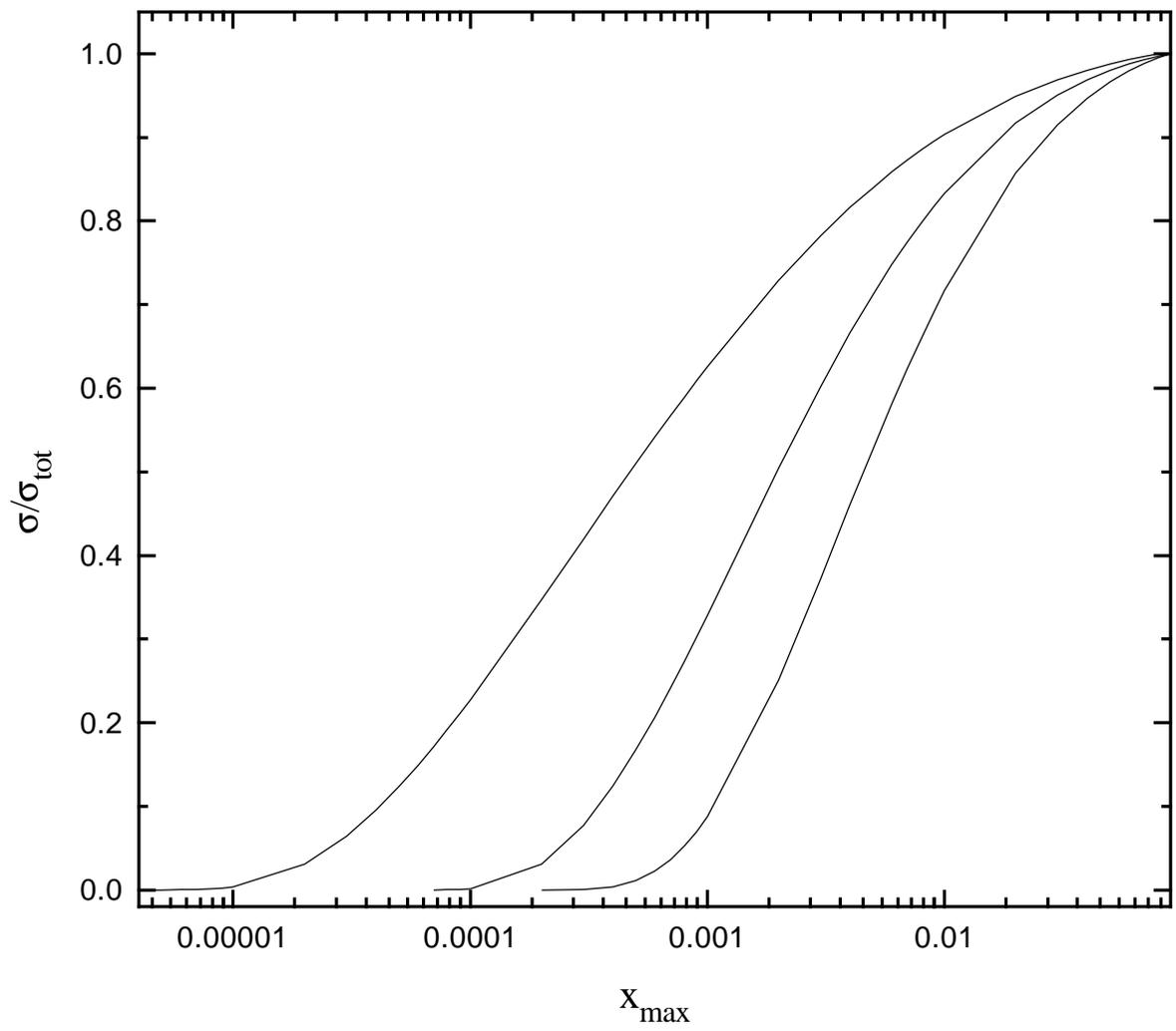
Fig.7.
The ration $\sigma/\sigma_{tot}$ as a function of $x_{\rm max}$,
where $\sigma_{tot}$ is the total cross section
after integrating over all $x$ region.
From left to right, the three curves correspond to three
lower bounds of transverse momentum, $k_{T{\rm min}}=0,~5,~10~GeV$
respectively.

\end{document}